\documentclass[aps,prb,reprint,superscriptaddress]{revtex4-1}

\usepackage{amsmath}
\usepackage{graphicx}
\usepackage{gensymb}
\usepackage{threeparttable}
\usepackage{multirow}
\usepackage[utf8]{inputenc}
\DeclareUnicodeCharacter{2212}{-}
\usepackage[T1]{fontenc}

\begin{document}


\title{Effects of thermal, elastic, and surface properties on the stability of SiC polytypes}

\author{Senja Ramakers}
\email[]{senja.ramakers@de.bosch.com}

\affiliation{Corporate Sector Research and Advance Engineering, Robert Bosch GmbH, Robert-Bosch-Campus 1, 71272 Renningen, Germany}
\affiliation{Interdisciplinary Centre for Advanced Materials Simulation, Ruhr-Universit\"at Bochum, Universit\"atsstra\ss e 150, 44801 Bochum Germany}
\author{Anika Marusczyk}
\affiliation{Corporate Sector Research and Advance Engineering, Robert Bosch GmbH, Robert-Bosch-Campus 1, 71272 Renningen, Germany}
\author{Maximilian Amsler}
\affiliation{Corporate Sector Research and Advance Engineering, Robert Bosch GmbH, Robert-Bosch-Campus 1, 71272 Renningen, Germany}
\author{Thomas Eckl}
\affiliation{Corporate Sector Research and Advance Engineering, Robert Bosch GmbH, Robert-Bosch-Campus 1, 71272 Renningen, Germany}
\author{Matous Mrovec}
\affiliation{Interdisciplinary Centre for Advanced Materials Simulation, Ruhr-Universit\"at Bochum, Universit\"atsstra\ss e 150, 44801 Bochum Germany}
\author{Thomas Hammerschmidt}
\affiliation{Interdisciplinary Centre for Advanced Materials Simulation, Ruhr-Universit\"at Bochum, Universit\"atsstra\ss e 150, 44801 Bochum Germany}
\author{Ralf Drautz}
\affiliation{Interdisciplinary Centre for Advanced Materials Simulation, Ruhr-Universit\"at Bochum, Universit\"atsstra\ss e 150, 44801 Bochum Germany}

\date{\today}

\begin{abstract}
SiC polytypes have been studied for decades, both experimentally and with atomistic simulations, yet no consensus has been reached on the factors that determine their stability and growth. Proposed governing factors are temperature-dependent differences in the bulk energy, biaxial strain induced through point defects, and surface properties. In this work, we investigate the thermodynamic stability of the 3C, 2H, 4H, and 6H polytypes with density functional theory (DFT) calculations. The small differences of the bulk energies between the polytypes can lead to intricate changes in their energetic ordering depending on the computational method. Therefore, we employ and compare various DFT-codes: VASP, CP2K, and FHI-aims; exchange-correlation functionals: LDA, PBE, PBEsol, PW91, HSE06, SCAN, and RTPSS; and nine different van der Waals (vdW) corrections. At $T=0$~K, 4H-SiC is marginally more stable than 3C-SiC, and the stability further increases with temperature by including entropic effects from lattice vibrations. Neither the most advanced vdW corrections nor strain on the lattice have a significant effect on the relative polytype stability. We further investigate the energies of the (0001) polytype surfaces that are commonly exposed during epitaxial growth. For Si-terminated surfaces, we find 3C-SiC to be significantly more stable than 4H-SiC. We conclude that the difference in surface energy is likely the driving force for 3C-nucleation, whereas the difference in the bulk thermodynamic stability slightly favors the 4H and 6H polytypes. In order to describe the polytype stability during crystal growth correctly, it is thus crucial to take into account both of these effects. 
\end{abstract}

\maketitle

\section{Introduction}\label{sec:intro}
Silicon carbide (SiC) is a wide bandgap IV-IV semiconductor with a polymorph-dependent gap between 2.3 and 3.3 eV \cite{kimoto_fundamentals_2014}, and has been touted to pave the route towards a new generation of semiconductor devices. It exhibits a plethora of advantages over conventional Si-based technologies, including a high electrical breakdown field of 2.5 MV/cm \cite{kaminski_sic_2014}, up to ten times higher than that of Si \cite{kimoto_material_2015}, which enables a reduction in material use and lowers the on-state resistance. SiC also has a high thermal stability, which allows for operations at high temperature with little to no degradation \cite{kimoto_material_2015}. The commercial production of a SiC-based metal-oxide semiconductor field-effect transistor MOSFET, the most common and important semiconductor device, has already been achieved \cite{stevanovic_recent_2010}.

\begin{figure}[t]
	\includegraphics[width=0.85\columnwidth]{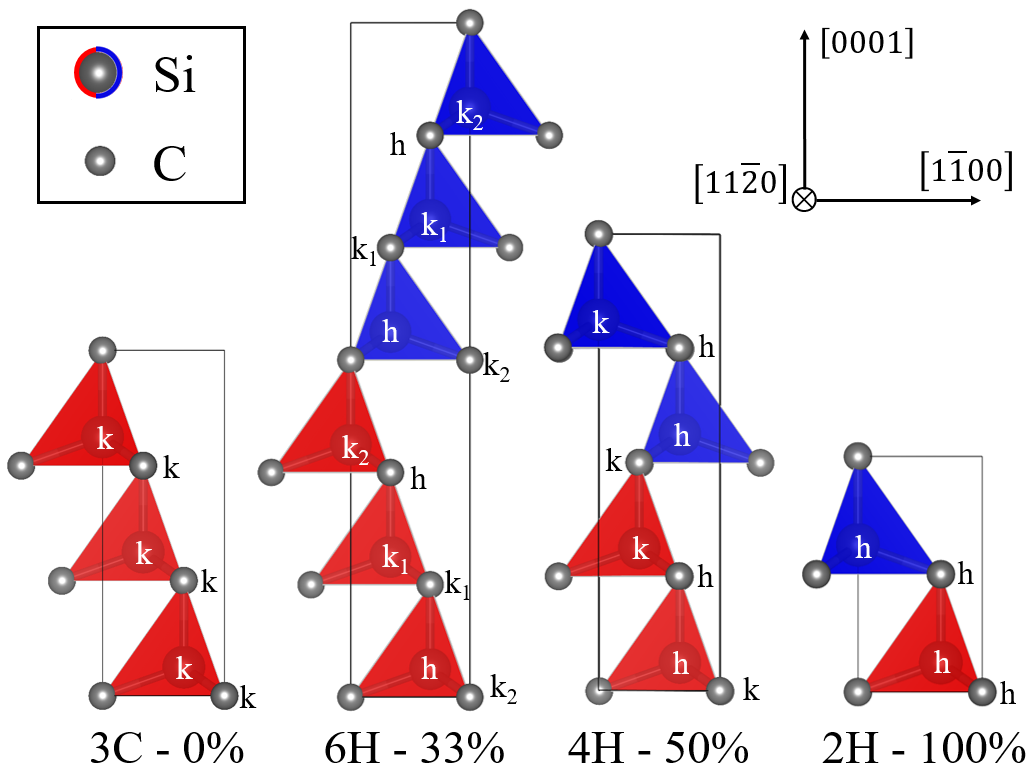}
	\centering
	\caption{The unit cells of the polytypes investigated in this work. The direction of the polyhedra, to the left (red) and right (blue), is directly related the hexagonal h-sites and cubic k-sites of the Si- and C-atoms; when the color from one layer to then next remains unchanged it indicates a k-site, whereas when the color changes it indicates a h-site.}\label{fig:schema}
\end{figure}

One of the biggest challenges in SiC processing is the growth of low-defect crystals due to the competing polymorphs that can form during synthesis. There exist over 200 SiC polymorphs, called polytypes, which crystallographically only differ in the stacking order of the SiC-bilayers. The most relevant and well-studied polytypes are the cubic 3C and the hexagonal 2H, 4H, and 6H polytypes. Their crystal structures are shown in Fig.~\ref{fig:schema} and their lattice parameters are listed in Table~\ref{tab:polytypes}. The difference in stacking patterns can be best visualized by viewing along the direction of the tetrahedral bonds for each bilayer in relation to the other layers. In Fig.~\ref{fig:schema}, when a tetrahedron is leaning towards the same direction as the lower bilayer (the same color), it is a cubic k-site, whereas when it is rotated by 180\degree\, (change in color), it is a hexagonal h-site. The polytype's hexagonality is defined by the relative fraction of such h-sites: $N_{\text{h}}/(N_{\text{h}}+N_{\text{k}})$. The 3C, 4H, and 6H polytypes have all been observed at similar growth conditions, complicating monocrystalline growth of the desired polytype \cite{stein_influence_1992, yakimova_polytype_2000, kimoto_bulk_2016}.

In the past decades, enormous advancements have been made in the fabrication of commercial substrate and low-defect epitaxial layer growth technology \cite{kimoto_bulk_2016, wada_extensive_2019, zhao_growth_2020}. The industry standard for SiC epitaxy is growth on 4H-SiC 4\degree off-axis Si-face substrates \cite{tsuchida_formation_2009, kimoto_bulk_2016} at process temperatures of around 1800 K. The formation of defects such as stacking faults \cite{yamashita_characterization_2015} and triangular defects \cite{guo_understanding_2017, yamashita_characterization_2018}, however, remain a major issue since they limit performance, cause leakage currents, lower the breakdown voltage and increase the on-state resistance\cite{kimoto_performance_1999, berechman_electrical_2009, lu_triangular_2018}. Such defects consist of a foreign polytype, often 3C, grown into the epitaxial layer during step-flow growth \cite{hallin_origin_1997, guo_understanding_2017, konstantinov_mechanism_1997, dudley_influence_1998, tsuchida_formation_2009, yamashita_characterization_2015, yamashita_characterization_2018}. Their origin has been attributed to different effects, like step-bunching \cite{kong_chemical_1988, ueda_crystal_1990, kimoto_surface_1994, syvajarvi_step-bunching_2002}, downfall particles \cite{guo_understanding_2017}, and substrate defects \cite{hallin_origin_1997, powell_controlled_1991, berechman_electrical_2009}. Yet, no conclusive consensus regarding the detailed mechanism of defect nucleation has been reached, a crucial requirement to develop advanced strategies to mitigate defect formation.

The fact that foreign polytypes have been found in 4H-SiC homoepitaxy \cite{heine_preference_1991, chien_terrace_1994, kimoto_bulk_2016, yakimova_polytype_2000, inomata_thermal_1969} demonstrates the importance of the thermodynamic stability and other influencing factors on the polytype specific growth. The surface termination of the substrate strongly correlates with the grown polytype \cite{stein_influence_1992}. In addition, thermal effects \cite{yakimova_polytype_2000, kimoto_bulk_2016, vasiliauskas_polytype_2014, arora_polytype_2020} and the influence of the gas composition and supersaturation \cite{fissel_thermodynamic_2000,  kakimoto_thermodynamic_2011, schmitt_polytype_2008, maltsev_4h-sic_1996} have been studied. Sublimation growth experiments have shown that 3C occurs at low temperatures, and 4H and 6H are only observed at high temperatures \cite{yakimova_polytype_2000, boulle_quantitative_2010, heine_preference_1991, inomata_thermal_1969}. 

Over the years, atomistic simulations have been used to study the phase stability, surface energetics, and growth kinetics of SiC polytypes. Many earlier \textit{ab initio} studies \cite{cheng_confirmation_1987, park_structural_1994, kackell_electronic_1994, karch_ab_1994,  limpijumnong_total_1998, jiang_ab_2002, bernstein_tight-binding_2005, konstantinova_ab_2008, mercier_role_2012, ito_theoretical_2011-1, ito_simple_2013, kawanishi_effect_2016} have reported on the thermodynamic stability using various methods \cite{troullier_efficient_1991, bachelet_pseudopotentials_1982, lin_ab_2013, vanderbilt_soft_1990, andersen_linear_1975, wei_local-density-functional_1985, tsakalakos_new_1984}, and we compiled a list of these results in Table~I of the SM \cite{SM}. In contrast to experimental findings, those reports predict that at $T=0$~K the 4H and 6H polytypes are slightly more stable (i.e., a few meV/SiC) than 3C, which indicates that the formation of 3C inclusions cannot be explained by such 0~K bulk energies alone. 

Several other determining factors that govern polytype stability and growth have been proposed in the literature. Scalise et al. \cite{scalise_temperature-dependent_2019} highlights the importance of employing an empirical Van der Waals dispersion correction \cite{grimme_semiempirical_2006, kawanishi_effect_2016} and considering the effect of lattice vibrations, which leads a polytype ordering consistent with temperature-dependent experimental data. Mercier and Nishizawa \cite{mercier_role_2012} showed that the 3C Si-terminated surface is more stable than the hexagonal surfaces, which could explain the high stability of 3C. Kang et al. \cite{kang_governing_2014} modeled the growth with a modified embedded atom method (MEAM) interatomic potential and reported that external conditions strongly influence the formation of point defects, which induce biaxial strain and could change the relative polytype stability. In view of all these possible mechanism at play, it remains a challenge to produce a model that reflects the correct materials behavior.

The goal of our work is two-fold. First, we benchmark the performance of different DFT methods by considering the influence of the exchange-correlation (XC) functional, pseudopotentials, basis-sets, and vdW corrections on various thermodynamic properties. Second, we study the energetics by comparing the bulk, thermal, elastic and surfaces of the most relevant SiC polytypes. 

\begin{table}[t]\caption{Overview of the crystal structures of SiC polytypes. For each polytype, the Ramsdell notation \cite{ramsdell_studies_1947}, bilayer stacking, ordering of $h$ and $k$ sites, the hexagonality $h/(h+k)$, and lattice constants $a$ and $c$ from Ref. \onlinecite{levinshtein_properties_2001} (3C, 6H, 4H) and from Ref. \onlinecite{harris_properties_1995} (2H) are shown.}\label{tab:polytypes}
	\setlength{\tabcolsep}{6.3pt}
	\begin{tabular}{lllll}
		\hline
		Polytype & Stacking       	   & Hexagonality & $a$ [\AA] & $c$ [\AA] \\ \hline
		3C  &     \footnotesize{(k)$_\infty$}  & 0            &         4.3596             &          4.3596             \\
		6H       &  \footnotesize{(k$_2$k$_1$h)$_\infty$} & 1/3 (33\%)   &         3.0805           &        15.1151              \\
		4H       &  \footnotesize{(kh)$_\infty$}  & 1/2 (50\%)   &            3.0798         &           10.0820           \\
		2H       & \footnotesize{(h)$_\infty$}   & 1 (100\%)    &            3.0763         &       5.0480            \\ \hline
	\end{tabular}
\end{table} 

\section{Computational methods}\label{sec:comp}
In this work, different flavors of density functional theory (DFT) \cite{hohenberg_inhomogeneous_1964, kohn_self-consistent_1965} are employed to investigate the thermodynamic properties of SiC. Three implementations of DFT are compared: the Vienna Ab Initio Simulation Package (VASP) \cite{kresse_ab_1993, kresse_ab_1994, kresse_efficiency_1996, kresse_efficient_1996}, the CP2K \cite{hutter_<span_2014, kuhne_cp2k:_2020} code, and the all-electron code FHI-aims \cite{blum_ab_2009, havu_efficient_2009}. 

VASP and CP2K both only treat the valence electrons explicitly and employ either the projector augmented wave (PAW) formalism \cite{blochl_improved_1994, kresse_ultrasoft_1999} or pseudopotentials, respectively, to represent the core states. For both methods, the four valence electrons, which participate in forming the covalent bonds in SiC, are treated as valence states. We employ the latest PAW-GW potentials (2015) for VASP with improved reproducibility \cite{lejaeghere_reproducibility_2016}. CP2K implements the Gaussian plane-wave (GPW) \cite{lippert_gaussian_1999} method and we employ the norm-conserving Goedecker-Teter-Hutter (GTH) pseudopotentials\cite{goedecker_separable_1996}.

Within CP2K, we compare the double-zeta valence short-range (DZVP-SR) basis-set to the triple-zeta valence (TZV2PX) basis-set, which stem both from the MOLOPT series \cite{vandevondele_gaussian_2007}. The number of Gaussian functions to describe the valence part of the wavefunction increases with the amount of valence splitting. The DZVP-SR basis employs four and five Gaussian functions, for Si and C respectively, whereas the TZV2PX basis employs six and seven. Moreover, the TZV2PX basis has two sets of polarization functions instead of one in the DZVP-SR.

The all-electron code FHI-aims employs the linear combination of atomic orbitals (LCAO) basis, with numerical atomic orbitals (NAO) to represent the wavefunction \cite{blum_ab_2009, havu_efficient_2009}.

In addition to the basis-sets and pseudopotentials, we investigate the influence of the XC functional on the thermodynamic and geometrical properties of SiC with the VASP-PAW method. We employ the local density approximation (LDA) \cite{perdew_self-interaction_1981}, the generalized gradient approximation (GGA) \cite{perdew_accurate_1992}, one hybrid functional, and meta-GGAs \cite{bohm_collective_1951, pines_collective_1952, bohm_collective_1953}. For LDA, we use the Ceperley-Alder (CA) functional \cite{ceperley_ground_1980}, while for GGA, we compare the Perdew-Wang (PW91) \cite{perdew_atoms_1992}, the Perdew-Burke-Ernzerhof (PBE) \cite{perdew_generalized_1996}, and the Perdew-Burke-Ernzerhof for solids and surfaces (PBEsol) \cite{perdew_restoring_2008} parametrizations. LDA is known to exhibit overbinding (i.e. prediction of lattice parameters smaller than experimental values), whereas the GGAs PW91 and PBE show underbinding. PBEsol was specifically parametrized to resolve these issues and provide lattice constants in better agreement with experiments. Furthermore, we employ the Heyd-Scuseria-Ernzerhof 06 (HSE06) \cite{heyd_hybrid_2003, krukau_influence_2006} hybrid functional, and two meta-GGA functionals, the revised Tao-Perdew-Staroverov-Scuseria (RTPSS) \cite{sun_self-consistent_2011} and the Strongly Constrained and Appropriately Normed (SCAN) \cite{sun_strongly_2015, sun_accurate_2016} functional. 

For each polytype we evaluate the bulk and (0001) surface properties. For the bulk calculations, we fully relax the lattice parameters and atomic coordinates with the conjugate gradient algorithm, making sure that the Pulay forces are well converged \cite{pulay_convergence_1980}. For the hybrid HSE06 functional, we took the PBE lattice and did not perform ionic or cell relaxations to reduce computational costs. The surface structures are created with the obtained relaxed lattice constants. For the surface calculations, we follow the approach of Ref. \onlinecite{mercier_role_2012} by including 7 SiC-bilayers and 12 $\text{\AA}$ vacuum along the surface normal. We relax the atomic positions of the upper four bilayers, while keeping the bottom three bilayers and the lattice parameters fixed. All calculations are carried out with periodic boundary conditions. We compare VASP, CP2K and FHI-aims with the commonly used PBE functional.

Well-converged total energies are essential because the energy differences between the various polytypes are very small, of the order of 1 meV/SiC. We adopt the same convergence criterion as in Ref. \onlinecite{de_waele_error_2016}, with an self-consistent field (SCF) tolerance of $10^{-8}$ eV and structural relaxation are performed until the change of the total energy is smaller than $10^{-6}$ $\text{eV}$ between two ionic steps. The convergence of the $k$-mesh and plane wave cut-off energy are shown in Fig.~1 of the SM \cite{SM}. The bulk calculations are converged to an energy change of \textless0.1 meV/SiC. We employ a $\Gamma$-point centered $k$-mesh. The k-mesh values are converged for the cubic 3C cell and adjusted for the cell sizes of the different polytypes. We choose a k-point density of 4096 k-points per reciprocal atom, which results in the values shown in Table~\ref{tab:k-points}. For the surface calculations, we create $1\times1$ 7-bilayer cells and employ a k-mesh of $13 \times 13 \times1$. Within VASP, a very high cut-off of 1500 eV is required to reach the convergence criteria of \textless0.1 meV/SiC and to obtain smooth energy-strain curves. Cut-off energies of 850 and 1200 eV are sufficiently accurate for the phonon and surface calculations, respectively, because the energy differences between the polytypes are much larger for these attributes. 

Besides the conventional DFT methods, we compare seven vdW energy correction methods. We employ three versions of the pairwise DFT-D$n$ methods by Grimme et al.: the DFT-D2 \cite{grimme_semiempirical_2006}, DFT-D3 \cite{grimme_effect_2011}, and DFT-D3/BJ, which is a version that includes Becke and Johnson damping \cite{becke_density-functional_2005, grimme_effect_2011}. Next, we consider two additional pairwise methods: the Tkatchenko-Scheffler (vdW-TS) \cite{tkatchenko_accurate_2009} method with and without iterative Hirshfeld (HI) partitioning \cite{bultinck_critical_2007, bucko_improved_2013, bucko_extending_2014}. Lastly, the many-body dispersion (vdW-MBD) method is the only correction method considered that goes beyond pair-wise interactions \cite{tkatchenko_accurate_2012, ambrosetti_long-range_2014}.

In addition, we employ three vdW XC functional, the vdW-DF2 functional \cite{dion_van_2004, langreth_van_2005, roman-perez_efficient_2009}, and the optimized optPBE-vdW and optB88-vdW functionals \cite{klimes_chemical_2010, klimes_van_2011}.

The thermal properties of SiC have been evaluated by Scalice et al. \cite{scalise_temperature-dependent_2019} We follow their approach and, in addition, assess the influence of the XC functional by employing LDA and PBE. Thermal effects are included by evaluating the vibrational energy contributions to the Helmholtz free energy within the harmonic approximation, $F_{\mathrm{vib}}(T) = U_{\mathrm{vib}}-TS_{\mathrm{vib}}$, where $U_{\mathrm{vib}}|_{T=0\mathrm{ K}}$ is the zero-point energy (ZPE) and $TS_{\mathrm{vib}}$ the entropy contribution. Configurational entropy is not considered because in a highly ordered crystal like SiC it is negligible. The vibrational entropy of the electronic contribution is sufficiently small that it can be approximated by its $T=0$~K limit computed by conventional DFT \cite{bechstedt_polytypism_1997, scalise_temperature-dependent_2019}.

\begin{table}[t]\caption{$k$-mesh employed for each polytype used for all calculations with VASP, CP2K and FHI-aims.}\label{tab:k-points}
	\begin{tabular}{llllll}
		\hline
		\small{Polytype} & 3C                & 6H                   & 4H                  & 2H                           \\ \hline
		\small{k-mesh} & \footnotesize{$8 \times 8 \times8\quad$} & \footnotesize{$13 \times 13 \times3\quad$} & \footnotesize{$13\times13\times4\quad$} & \footnotesize{$13\times13\times8$}  \\ \hline
	\end{tabular}
\end{table}

To obtain the lattice's vibrational free energy, we first compute the interatomic force constants with density functional perturbation theory (DFPT) \cite{giannozzi_ab_1991, gonze_dynamical_1997}. Thereafter, the dynamical matrices, phonon dispersion, and thermal properties are determined using Phonopy \cite{togo_first_2015}. To obtain the interatomic force constants we perform DFPT calculation with VASP at $\Gamma$-point in supercells of dimensions $3 \times 3 \times 1$  for 4H and 6H,  $3 \times 3 \times 2$ for 2H, and $2 \times 2 \times 2$ for 3C.  

We also compute the elastic properties with the DFPT method as implemented in VASP. The stiffness tensor is determined by performing six finite distortions of the lattice \cite{le_page_symmetry-general_2002}. The elastic constants are derived from the stress-strain relationship according to Hooke's law $\sigma_i =C_{ij}\epsilon_j$, where $C_{ij}$ denote the elastic constants using Voigt notation, $\sigma_i$ the stress, and $\epsilon_j$ the strain. The bulk ($K$) and shear ($G$) moduli are calculated using the Voigt-Reuss-Hill (VRH) average \cite{man_simple_2011}. First, the Voigt upper bound and Reuss lower bound are determined. These two bounds are simply averaged to obtain $K_{\text{VRH}}$ and $G_{\text{VRH}}$. The Poisson ratio can be derived as $\nu = (3K-2G)/(6K+2G)$, and the Young's modulus as $E = (9KG)/(3K+G)$. We keep the same supercells and computational settings as employed in the thermal properties calculations.

\section{Results and discussion}
\subsection{Polytype stability of the bulk}\label{ssec:bulk}
We compute the relative thermodynamic stability of the SiC polytypes at $T=0$~K by comparing the internal energy of 3C, 2H, 4H, and 6H. In this Section, we first compare results of the LDA, PBE, PBEsol, and PW91, the hybrid HSE06, and the meta-GGA SCAN and RTPSS functionals within the VASP-PAW method. Next, the different \textit{ab-initio} methods VASP-PAW, CP2K-GPW, and FHI-aims-LCOA are compared. Lastly, we study the influence of the chosen basis-set with CP2K-GPW and FHI-aims-LCAO. 

\begin{figure}[t]
	\includegraphics[width=\columnwidth]{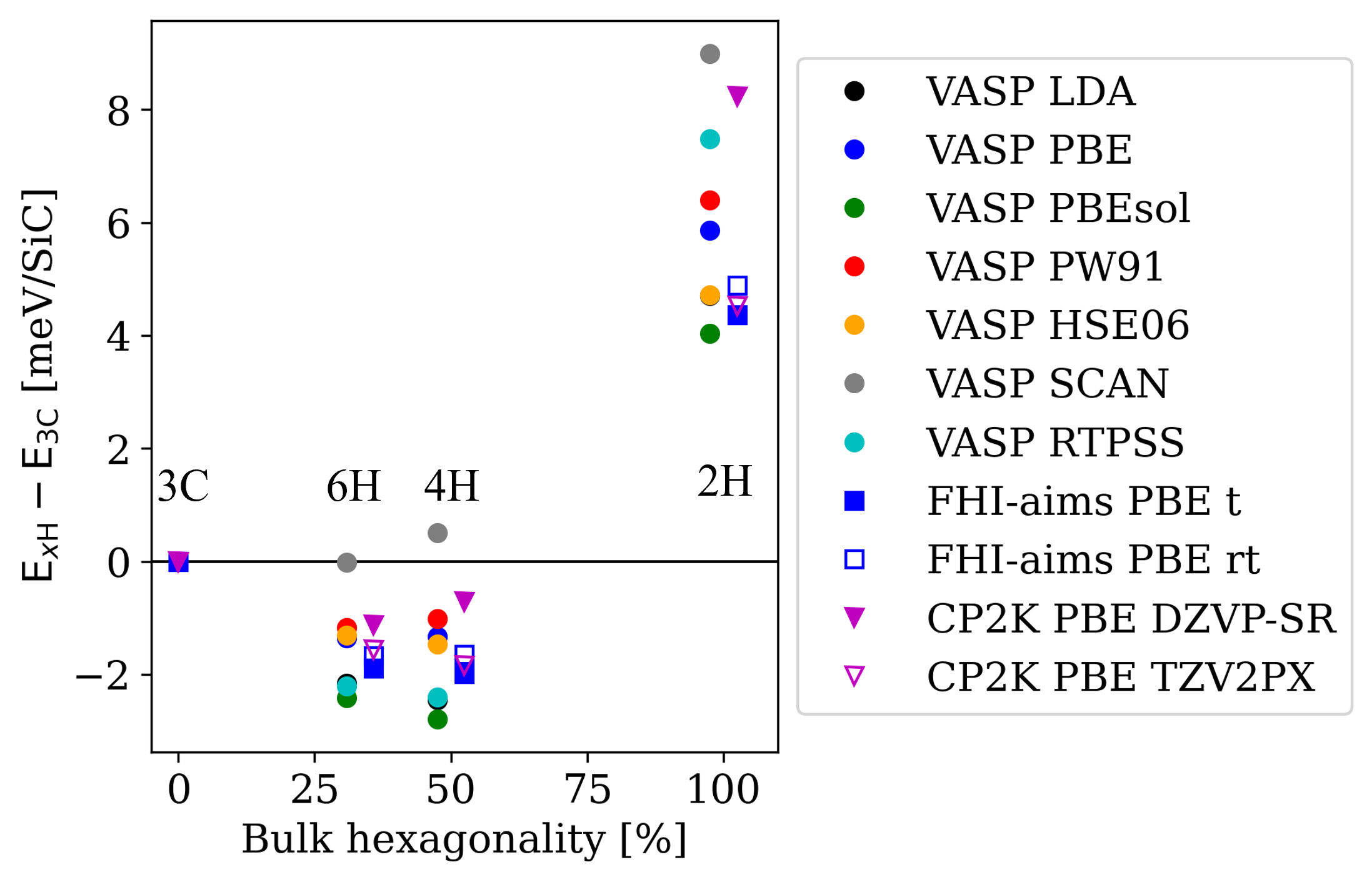}
	\centering
	\caption{Energy differences of the relaxed 3C, 6H, 4H, and 2H structures as a function of hexagonality for different DFT codes, XC functionals, and basis-sets. FHI-aims t (rt) stands for tight (really tight). To improve visualization, we shifted the VASP data to the left and the FHI-aims and CP2K data to the right.}\label{fig:bulk}
\end{figure}

An overview of the polytype's stacking and hexagonality is given in Table~\ref{tab:polytypes}. The $x$H polytypes are compared to 3C using $\Delta  E_{x\text{H}}=E_{x\text{H}}-E_{\text{3C}}$, for $x=2,4,6$. When $\Delta E_{\text{xH}}$/SiC-pair is positive (negative), the cubic (hexagonal) polytype is more stable. The relative energies versus hexagonality are plotted in Fig.~\ref{fig:bulk} and listed in Table~II of the SM \cite{SM}. The results are obtained for the relaxed lattice parameter for each computational method. The effect of the lattice parameter on the energy differences is negligible and does not change the relative ordering. We included the same study with fixed lattice parameters from Refs. \onlinecite{levinshtein_properties_2001, harris_properties_1995} in Fig.~2 of the SM \cite{SM}. 

Contradictory to experimental findings, conventional DFT predicts that the 4H and 6H polytypes are slightly more stable than 3C. All methods besides the meta-GGA SCAN functional show a similar relative polytype stability, where the 4H and 6H polytypes are -1 to -5 meV/SiC more stable than 3C. For each method, the difference between $\Delta E_{4\text{H}}$ and $\Delta E_{6\text{H}}$ is \textless0.25 meV/SiC. The relative stability of 4H and 6H differs with the employed method. 2H  is the least stable polytype by 1 to 8 meV/SiC compared to 3C. These results correspond well to other DFT computations in the literature \cite{cheng_confirmation_1987, park_structural_1994, kackell_electronic_1994, karch_ab_1994,  limpijumnong_total_1998, jiang_ab_2002, konstantinova_ab_2008, mercier_role_2012, ito_simple_2013, kawanishi_effect_2016}.  

Next, we investigate the influence of the XC functional on the thermodynamic stability with VASP. The LDA functional, which only takes into account the local electron density, shows an energy difference between 3C and 4H/6H of around -2 meV/SiC. The GGA functionals PBE and PW91, which take into account the first derivative of the electron density, show smaller energy difference of -1 meV/SiC. We further employ the HSE06 hybrid functional, which incorporates a fraction of the exact exchange energy based on the geometries obtained with PBE. For 4H and 6H, the same energy differences are observed as PBE/PW91, showing good agreement with the semilocal DFT functionals. PBEsol is a GGA functional that was parameterized to improve the description of solids, for which the relative energies are similar to LDA. The meta-GGA functionals SCAN and RTPSS include the second derivative of the electron density. RTPSS reproduces similar results as LDA/PBEsol. In contrast, SCAN predicts 3C, 4H, and 6H to have close to identical energy values. For all methods, the 2H polytype is clearly unstable compared to all other polytypes, which agrees well with the fact that this polytype has not been observed in growth experiments.

To study the reproducibility of our findings, we compare the CP2K-GPW and FHI-aims-LCAO methods to VASP-PAW while employing the PBE functional. Within CP2K, we investigate two basis-sets which differ in their size: DZVP-SR and TZV2PX. The results for the larger TZV2PX basis-set are expected to be more accurate. Fig.~\ref{fig:bulk} shows that the differences between the two basis-sets are significant, especially for 2H. For 4H and 6H, the CP2K results resemble the energies of VASP-PBE. Due to the small energy difference between these two polytypes, the relative stability of 4H and 6H even changes between the two basis-sets. For 2H, the DZVP-SR basis-set shows a large discrepancy of up to 4 meV compared to the other methods VASP-PAW-PBE. For the all-electron code FHI-aims, the difference between the "tight" and "really tight" basis-sets is insignificant. Generally, FHI-aims produces results consistent with VASP and even more so with CP2K-TZV2PX.  

\begin{figure}[t]
	\includegraphics[width=\columnwidth]{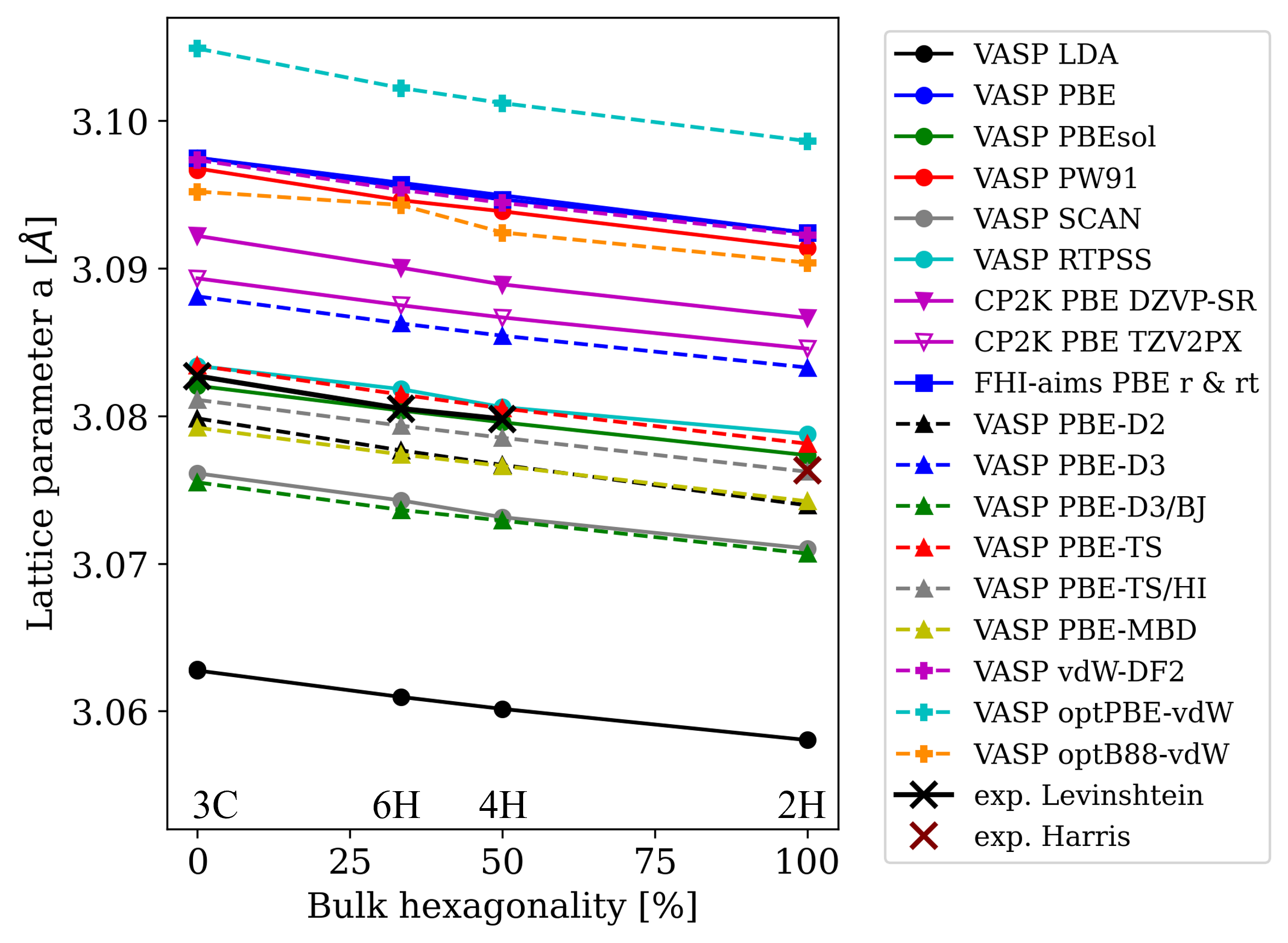}
	\centering
	\caption{Lattice constants $a$ versus bulk hexagonality for the conventional DFT and van der Waals corrections. FHI-aims t (rt) stands for tight (really tight). Experimental X-ray diffraction results of Levinshtein et al. \cite{levinshtein_properties_2001} and Harris et al. \cite{harris_properties_1995} are included.} \label{fig:lattice_a}
\end{figure}

In addition to the relative stability, we analyze the influence of the computational method on the relaxed lattice constants. In the hexagonal structures, the lattice constants $a$, $b$, $c$ correspond to the $[11\bar{2}0]$, $[1\bar{2}10]$ and  $[0001]$ directions, respectively. In Fig.~\ref{fig:lattice_a}, $a$ is plotted versus the polytype hexagonality. In order to compare the properties of the cubic and hexagonal cells, we map the cubic cell onto a hexagonal lattice. Then, the [111] direction in the cubic lattice aligns with the [0001] direction of the hexagonal lattice, which results in $a_{\text{H}}=a_{\text{C}}/\sqrt{2}$ and $c_{\text{H}}=\sqrt{3}c_{\text{C}}$. 

The value of $a$ decreases for increasing polytype hexagonality, a trend that has been verified with X-ray diffraction experiments \cite{levinshtein_properties_2001, stockmeier_lattice_2009}. For the non-vdW methods, we find strong overbinding for LDA, and to a lesser extent for SCAN. On the other hand,  PBE (within all DFT implementations) and PW91 show underbinding. The PBEsol and RTPSS functionals resolve this issue and predict values very close to experiments \cite{levinshtein_properties_2001, stockmeier_lattice_2009, harris_properties_1995}.

An analysis including the lattice constant along the stacking direction $c$ is shown in Fig.~\ref{fig:lattice_ac}. Since the number of SiC bilayers within one unit cell varies for the different polytypes, the ratio of the lattice constants $c/(2xa)$ is calculated, where $x=2,4,6$ for the $x$H polytypes. For 3C ($x=3$), this leads to the ideal ratio $c/(2xa)=\sqrt{2/3}$. The ratio $c/(2xa)$ increases linearly with hexagonality, which means that the higher the polytype hexagonality, the narrower and more elongated the lattice becomes. The visualization in Fig.~\ref{fig:lattice_ac} shows that almost all conventional functionals the ration of lattice constants well compared to experiments. Only CP2K-DZVP-SR shows a rather large value since it overpredicts the elongation and narrowing of the hexagonal cells. All values for the lattice parameters $a$ and $c$ are listed in Table~III of the SM \cite{SM}. 
 
\subsection{Van der Waals corrections}\label{ssec:bulk_vdw}
Long range van der Waals (vdW) dispersion forces are usually relevant in molecular systems and layered crystals but rather negligible in covalently bonded materials. However, results of Kawanishi and Mizoguchi \cite{kawanishi_effect_2016} suggest that the SiC polytype stability is highly dependent on vdW dispersions. They applied the DFT-D2 dispersion correction \cite{grimme_accurate_2004, grimme_semiempirical_2006} to the VASP-PBE method. In their calculations, the vdW contribution to the total energy was around 3\%, which changed the relative polytype stability drastically. The DFT-D2 method by Grimme et al. \cite{grimme_semiempirical_2006} is a semiempirical approach, which does not account for any effects of the chemical environment nor does it show the correct asymptotic behavior \cite{grimme_consistent_2010}. Generally, it is recommended to use the more recent DFT-D3 correction \cite{stohr_theory_2019}. 

In this section, we investigate the bulk properties with DFT-D2 and more advanced vdW dispersion methods. The differences in implementation of the vdW correction are discussed in Section~I of the SM \cite{SM} and in Ref. \cite{hermann_first-principles_2017}. We compare a total of nine vdW methods, which consist of either an added correction applied to the VAPS-PBE total energy or an independent XC functional that incorporates the vdW dispersion. 

\begin{figure}[t]
	\includegraphics[width=\columnwidth]{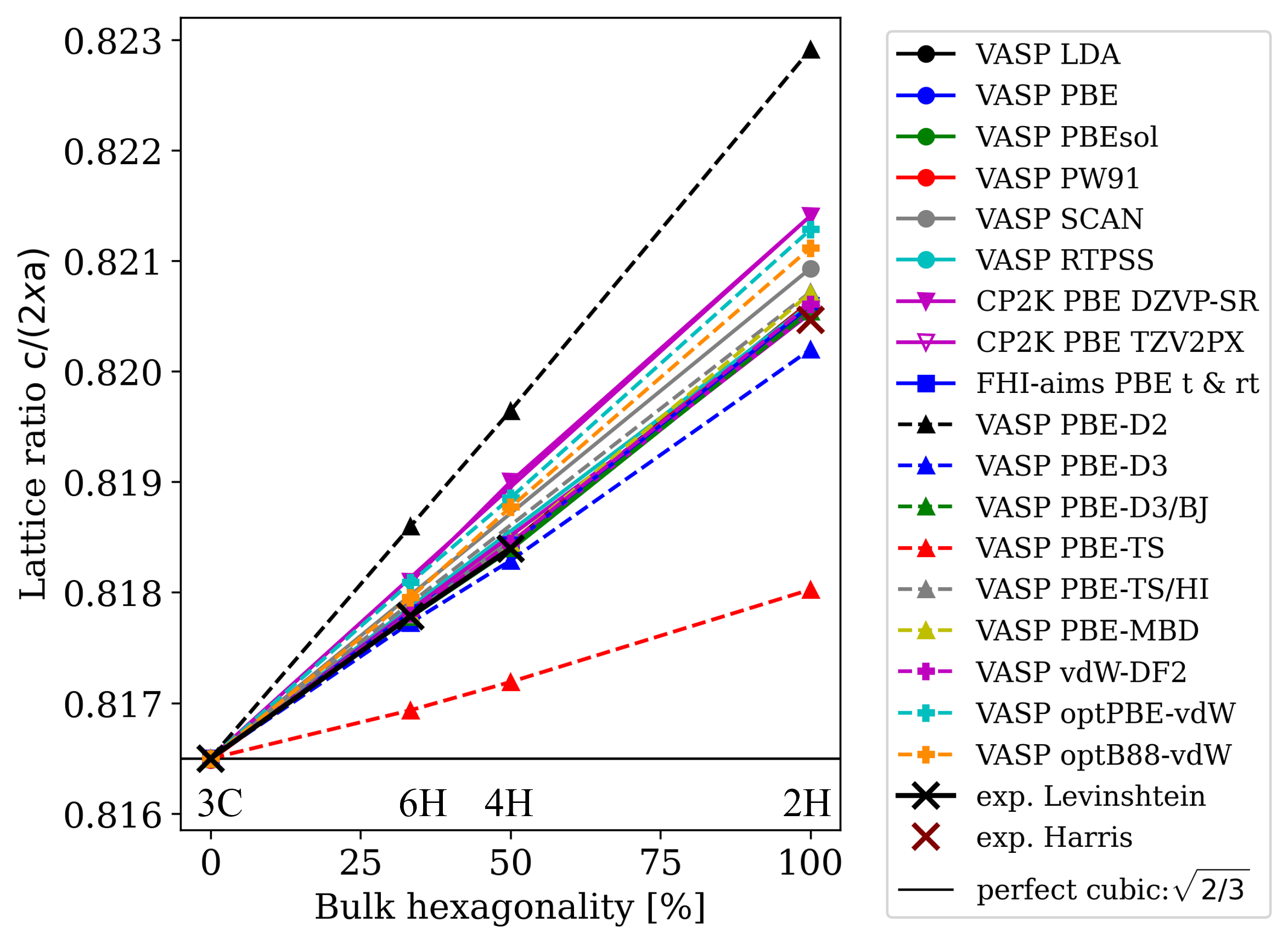}
	\centering
	\caption{Normalized ratio of lattice constants $c/(2xa)$ versus bulk hexagonality for the conventional DFT and vdW dispersion methods. FHI-aims t (rt) stands for tight (really tight). Experimental X-ray diffraction results of Levinshtein et al. \cite{levinshtein_properties_2001} and Harris et al. \cite{harris_properties_1995} are included.} \label{fig:lattice_ac}
\end{figure}

\begin{figure}[t]
	\includegraphics[width=\columnwidth]{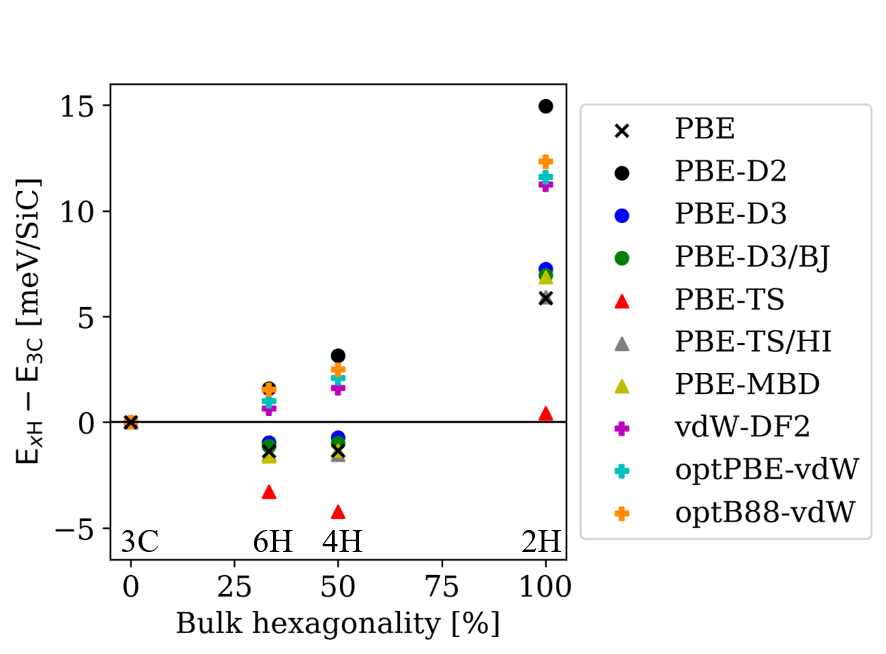}
	\centering
	\caption{Energy differences for 3C, 6H, 4H and 2H-SiC as a function of hexagonality, calculated either with VASP-PBE and a vdW correction or with a vdW XC functional.} \label{fig:bulk_vdw}
\end{figure}

The results for the polytype stability calculated by employing the various vdW dispersion methods are shown in Fig.~\ref{fig:bulk_vdw}. The VASP-PBE energies are shown as a reference. Clearly, the spread in energy differences is much larger compared to the results obtained with the conventional DFT methods shown in Fig.~\ref{fig:bulk}. We are able to exactly reproduce the results for the DFT-D2 method of Ref. \cite{kawanishi_effect_2016}. This method indeed strongly stabilizes 3C. However, calculations with the more advanced DFT-D3 method \cite{grimme_consistent_2010} have a completely different outcome; 3C is only slightly stabilized, but the relative stability remains unchanged compared to uncorrected PBE. An explanation could be that DFT-D2 only includes the 6th-order dispersion, whereas DFT-D3 also includes the 8th-order. Our results indicate that in DFT-D3, the 6th- and 8th-order cancel each other, and the effect of the relative stability is very small compared to DFT-D2. The addition of Becke and Johnson damping, DFT-D3/BJ \cite{becke_density-functional_2005, grimme_effect_2011}, does not have a significant effect on the DFT-D3 results. 

To obtain a better understanding of the effect of the DFT-D$n$ corrections, we also analyze the lattice constants. Fig.~\ref{fig:lattice_a} shows that all three methods correct the underbinding and overestimation of $a$ observed for PBE. The amount of correction differs notably though. The DFT-D2 predicted lattice parameter is closest to experiments among the DFT-D$n$ methods. For DFT-D3, the value of $a$ remains greater than the experimental one, whereas for DFT-D3/BJ the lattice parameter is substantially smaller. The lattice ratios, shown in Fig.~\ref{fig:lattice_ac}, are substantially different between the three DFT-D$n$ methods. DFT-D3 shows good agreement with experimental values, and the addition of BJ-damping results in a slight underestimation of the lattice ratio. A large overestimation is observed for DFT-D2, which indicates that this method might not be well suited for SiC.

In addition to the DFT-D$n$ methods, we evaluate various other vdW correction methods: vdW-TS \cite{tkatchenko_accurate_2009}, vdW-TS/HI \cite{bucko_improved_2013, bucko_extending_2014, bultinck_critical_2007}, and vdW-MBD \cite{tkatchenko_accurate_2012, ambrosetti_long-range_2014}. The vdW-TS method describes the polytype stability poorly by predicting 2H and 3C to be equally stable, which disagrees with the fact that 2H is not observed in growth experiments \cite{kimoto_bulk_2016}. Although it does predict the lattice constant $a$ quite accurately, the predicted lattice ratio is clearly underestimated compared to experimental values. The vdW-TS method is known to describe ionic solids poorly \cite{bucko_extending_2014}, which can be improved by  applying iterative Hirshfeld partitioning (vdw-TS/HI). Although SiC is far from an ionic solid, the Hirshfeld partitioning does improve both the relative energies and the lattice constants. Now, the effect on the relative polytype stability diminishes and the results behave like PBE. 

The most advanced and expensive correction method considered in this work is the vdW-MBD correction, which includes many-body vdW-interactions based on the random phase approximation correlation energy \cite{tkatchenko_accurate_2012, ambrosetti_long-range_2014}. Here again, the correction is negligible and we obtain the uncorrected PBE results. The lattice constant $a$ is slightly underestimated, but the ratio of lattice constants is predicted correctly. 

Lastly, three vdW XC functionals are evaluated. The vdW-DF2 \cite{dion_van_2004, langreth_van_2005, lee_higher-accuracy_2010}, optPBE-vdW \cite{klimes_chemical_2010, klimes_van_2011}, and optB88-vdW \cite{klimes_chemical_2010, klimes_van_2011} are all based on the same principle of including vdW dispersion in the functional directly. The methods differ in the exchange part of the functional but not the correlation part, which is where the vdW dispersion contributes. All three functionals predict 3C to be the most stable and differ very little from each other, the minor differences are likely caused by the exchange part of the functionals. For both the polytype stability and the structural optimization, the results are similar to the values obtained for DFT-D2. The lattice parameter $a$ is not improved with respect to the PBE functional. In the case of optPBE-vdW, $a$ is even more overestimated. For all three functionals, Fig~\ref{fig:lattice_ac} shows that the lattice ratio $c/(2xa)$ is overestimated compared to experiments. 

To conclude, we observe that the various vdW-methods investigated here differ significantly in outcome and the relative order of the polytype stability. The lattice ratio $c/(2xa)$ could serve as an indicator for how well a specific vdW method is suited to describe the SiC system. The methods which describe the lattice most accurately are vdW-D3(BJ), vdW-TS/HI, and vdW-MBD, which all predict that vdW-dispersion has an insignificant effect on the polytype stability. The PBEsol XC functional remains the most accurate method for geometrical properties. For this reason we limit ourselves to conventional, non-vdW methods throughout the rest of this work. In general, the $T=0$~K internal energies of 3C, 4H, and 6H are all within a range of merely a few meV/SiC, so that we have to consider other effects like the thermal, mechanical, and surface properties to explain experimental observations. 

\begin{figure}[t]
	\includegraphics[width=\columnwidth]{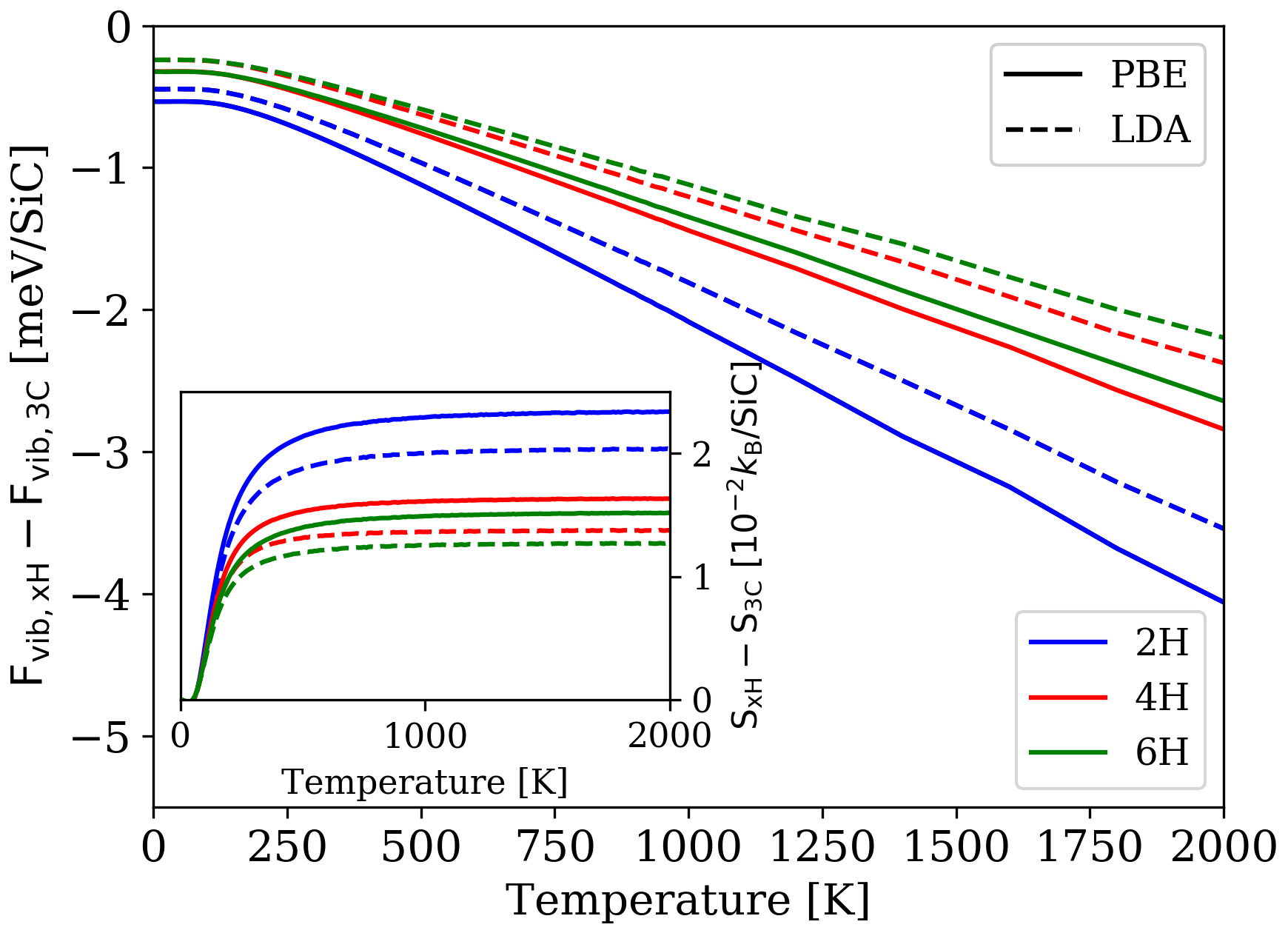}
	\centering
	\caption{The phonon vibrational free energy of the xH polytypes relative to 3C, calculated for the LDA and PBE XC functional in the harmonic approximation. The inset shows the difference in vibrational entropy.} \label{fig:thermal}
\end{figure}

\subsection{Temperature contributions}\label{ssec:results_thermal}
SiC crystal growth processes take place at very high temperatures. Sublimation growth requires temperatures as high as 2650 K, whereas the chemical vapor deposition (CVD) epitaxial growth occurs at 1800~K \cite{kimoto_bulk_2016}. 
Therefore, it is important to evaluate temperature contribution to the thermodynamic stability of the polytypes. Scalise et al. \cite{scalise_temperature-dependent_2019} assert that the temperature-dependent polytype specific growth (i.e., 3C at lower $T$, 4H and 6H at higher $T$) can be explained by taking into account this contribution. We follow their approach by calculating the phonon spectrum with DFPT \cite{giannozzi_ab_1991, gonze_dynamical_1997} as described in Section~\ref{sec:comp}. Since the static internal energy differences between the polytypes ($\Delta U_{0}$) vary for different computational methods, as discussed in Sections~\ref{ssec:bulk} and~\ref{ssec:bulk_vdw}, we  focus here on the vibrational contribution to the free energy $F_{\mathrm{vib}}(T) = U_{\mathrm{vib}}-TS_{\mathrm{vib}}$. Furthermore, we study the influence of the XC functional by comparing the LDA and PBE functionals.

In Fig.~\ref{fig:thermal}, the relative phonon vibrational contribution to the free energy is shown with respect to the 3C polytype. We find that $\Delta F_{\text{vib}}$ is negative and decreases with temperature for all hexagonal polytypes, which means that increased temperatures stabilize the hexagonal polytypes. Both XC functionals PBE and LDA show the same trend overall, but the stabilization of the hexagonal polytypes is slightly stronger for PBE.

\begin{figure}[t]
	\includegraphics[width=0.95\columnwidth]{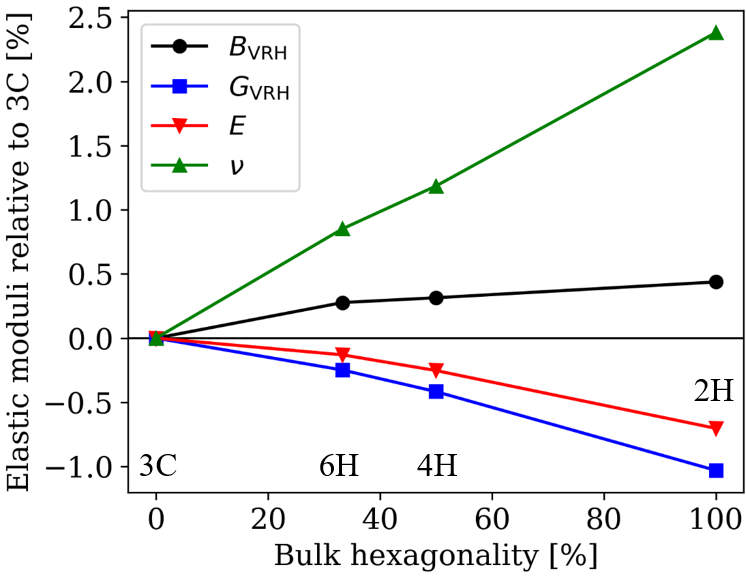}
	\centering
	\caption{The change in elastic moduli relative to 3C versus bulk hexagonality.} \label{fig:elastic_moduli}
\end{figure}

The zero-point energy differences ($\Delta U_{\text{vib}}|_{T=0\mathrm{K}}$) between 3C and the hexagonal polytypes is small, around -0.5 meV. The values of $\Delta S_{\text{vib}}$ among the polytypes is shown in the inset of Fig.~\ref{fig:thermal}. In the temperature range $T<300$ K, $\Delta S_{\text{vib}}$ increases steadily from zero to its maximum value. In this range, $\Delta F_{\text{vib}}$ remains close to $\Delta U_{\text{vib}}|_{T=0\mathrm{ K}}$. For $T>300$ K, $\Delta S_{\text{vib}}$ approaches a constant value, which results in $\Delta F_{\text{vib}}$ increasing linearly. The highest entropy is found for the 2H polytype. The 4H and 6H polytypes exhibit very similar behavior; the 4H polytype, with a higher hexagonality than 6H, shows a higher  $\Delta S_{\text{vib}}$ and lower $\Delta F_{\text{vib}}$. 

At the CVD processing temperature of around 1800~K, the phonon vibrational energy stabilizes the hexagonal polytypes 4H (6H) over 3C by about -3.0 (-2.8) meV/SiC. This temperature contribution is added to the static internal energy differences at $T=0$~K ($\Delta U_{0}$) to obtain the Helmoltz free energy differences $F=U_{0} +  U_{\text{vib}} - TS_{\mathrm{vib}}$. Scalise et al. \cite{scalise_temperature-dependent_2019} obtain similar results for the temperature contribution but take the $U_0$ value from the PBE-D2 vdW correction method. We have previously shown that the PBE-D2 correction does not well reproduce experimental lattice parameters and should therefore be omitted. Our findings of $U_0$ (in Sections~\ref{ssec:bulk} and ~\ref{ssec:bulk_vdw}) show that for conventional DFT methods and most vdW corrections 4H and 6H are equally or up to -5 meV/SiC more stable than 3C. The temperature contribution only further increases the thermodynamic stability of 4H and 6H.

To conclude, the thermodynamic stability of the bulk, including the vibrational free energy, cannot by itself explain experimental findings, in which 3C is observed at lower growth temperatures and 3C
inclusions pose severe problems \cite{guo_analysis_2014, berechman_electrical_2009}. Note, however, that we did not take into account any anharmonic effects, which become increasingly important at these high processing temperatures and might affect the energetic ordering.

\begin{figure}[t]
	\includegraphics[width=0.95\columnwidth]{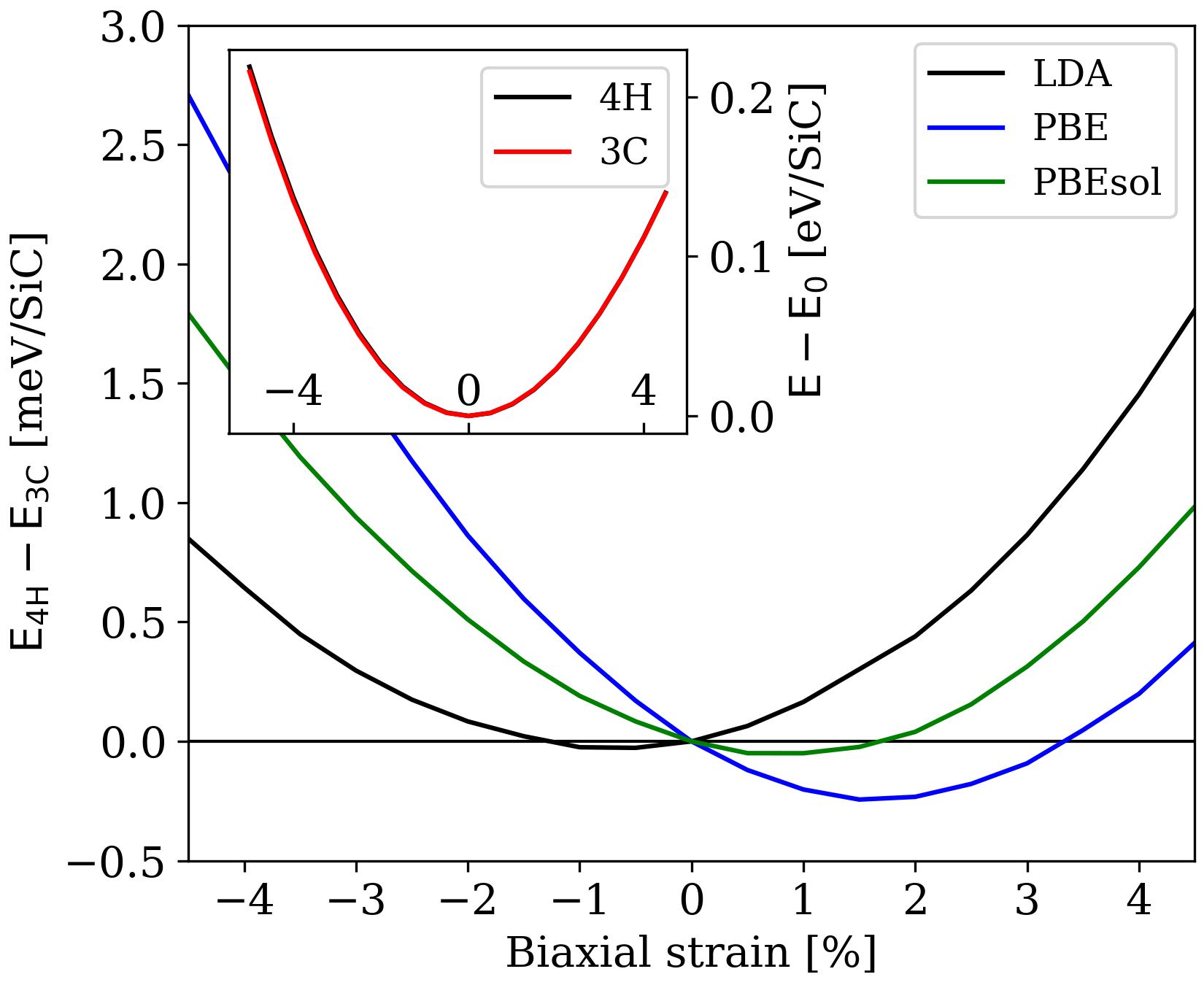}
	\centering
	\caption{Effects of (0001) biaxial strain on the relative stability of 3C and 4H-SiC at $T = 0$ K for three XC functionals. The inset shows the energy-strain curves for 3C and 4H. Elastic energy difference between 3C and 4H is plotted versus biaxial strain as a percentage of the lattice constant $a$, where a positive difference means 3C is more stable.} \label{fig:ev_curve}
\end{figure}   

\subsection{Elastic properties}\label{ssec:results_elastic}
\begin{table*}[t]\caption{Elastic constants ($C_{ij}$ in GPa), the Voigt (V), Reuss (R) and Voigt-Reuss-Hill (VRH) average of the bulk modulus ($B$ in GPa) and of the shear modulus ($G$ in GPa), Young's modulus ($E$ in GPa), Poisson’s ratio ($\nu$) calculated with the VASP-PBE method in the framework of DFPT. The results are compared to experimental values.}\label{tab:elastic_constants}
	\setlength{\tabcolsep}{8pt} 
	\begin{threeparttable}	
		\begin{tabular}{lllllllll}
			\hline 		   	& 3C	&      	& 6H    &		& 4H	&		& 2H 	&     \\ 
			& DFT	&  Exp.	& DFT	&  Exp.	& DFT	&  Exp. & DFT	&  Exp. \\ \hline
			$C_{11}$       	& 384.3 & 390\tnote{a}, 395\tnote{e}, 371\tnote{g}	& 485.2 & 501\tnote{d}	&487.8 & 501\tnote{d}	& 494.3& 	\\
			$C_{12}$      	& 127.9 & 142\tnote{a}, 123\tnote{e}, 146\tnote{g}	& 106.0 & 112\tnote{d}	&105.0 & 111\tnote{d}	&102.1 &	\\
			$C_{13}$       	& -		& - & 52.3  & 52\tnote{d}	& 51.9  & 52\tnote{d}	&50.7  &	\\
			$C_{33}$       	& -		& -	& 534.1 &  553\tnote{d}	& 533.4 & 553\tnote{d}	&533.5 &	\\
			$C_{44}$       	& 239.9 & 150\tnote{b}, 256\tnote{a}, 236\tnote{e},111\tnote{g}	& 160.1 &163\tnote{d} 	&157.9 & 163\tnote{d}	&151.3 &	\\
			$C_{66}$       	& - 	& -	& 189.6 & 	&191.4 &	&196.1 &	\\
			$B_{\text{V}}$ 	& 213.6	&  	& 213.0 &  	&214.0 & 	&214.3 &	\\
			$B_{\text{R}}$	 & 213.6&   & 212.9 & 	&214.0 & 	&214.3 &    \\
			$B_{\text{VRH}}$ & 213.6&   225\tnote{a}, 270\tnote{b}  & 213.0 & 	&214.0 & 	&214.3 & 223\tnote{c}  \\
			$G_{\text{V}}$   & 195.4&  	& 188.2	& 	&188.1 &  	&187.6 &    \\
			$G_{\text{R}}$ 	 & 178.0&   & 184.0 & 	&183.5 &  	&181.6 &    \\
			$G_{\text{VRH}}$ & 186.7& 192\tnote{c} 	& 186.1 & 	&185.8 & 	&184.6 &    \\
			$E$				 & 433.7&  448\tnote{c} & 432.8 & 450\tnote{f}	&432.3 & 450\tnote{f}	&430.4 &	\\
			$\nu$			 &0.1616&  0.267\tnote{a}, 0.168\tnote{c}	&0.1628 &	&0.1634&	&0.1653&	\\ \hline
		\end{tabular}
		\begin{tablenotes}\footnotesize
			\item[a] Ref. \onlinecite{feldman_phonon_1968}, 
			\item[b] Ref. \onlinecite{harrison_electronic_1981}
			\item[c] Ref. \onlinecite{carnahan_elastic_1968}
			\item[d] Ref. \onlinecite{kamitani_elastic_1997}
			\item[e] Ref. \onlinecite{djemia_elastic_2004}
			\item[f] Ref. \onlinecite{wolfenden_measurements_1996}
			\item[g] Ref. \onlinecite{pestka_measurement_2008}, 3C-SiC thin film
		\end{tablenotes}
	\end{threeparttable}
\end{table*}
During synthesis, the wafers are exposed to thermal stress, which can influence the stability of the epitaxially grown polytypes. In addition, point defects and doping could lead to a deformation of the lattice depending on the concentration as suggested by Kang et al. \cite{kang_governing_2014}. In this Section, we investigate the difference in elastic properties between the polytypes and how it influences the thermodynamic stability. We compute the stiffness tensor, consisting of the elastic constants $C_{ij}$, with DFPT. From these, the elastic moduli are determined as described in Section~\ref{sec:comp}. Next, we study the effect of biaxial strain on the polytype stability. 

An overview of the elastic constants from VASP-PBE DFPT and experimental reference data (whenever available) is shown in Table~\ref{tab:elastic_constants}. For 3C-SiC, in particular, multiple experimental reference values have been reported in the literature, some of which however substantially deviate from each other. Nevertheless, we compare the average results to our computed values and observe that both the elastic constant and moduli are underestimated by up to 10\%. The elastic constants computed for 4H and 6H are in good agreement with experimental results from Brillouin scattering \cite{kamitani_elastic_1997}.

While the components $C_{ij}$ of the cubic and hexagonal polytypes are not directly comparable with each other due to their different lattices,  it is possible to evaluate and compare the averaged bulk and shear moduli. While for the bulk modulus $B$ the Voigt upper and Reuss lower bound are equal, this is not the case for the shear modulus $G$. We determined the Young's moduli $E$ and the Poisson ratio $\nu$ based on $B_{\text{VRH}}$ and $G_{\text{VRH}}$. 

The differences between the polytypes are visualized in Fig.~\ref{fig:elastic_moduli}, where the change in the moduli relative to 3C is shown. The elastic moduli $B$ and $\nu$ increase, whereas $G$ and $E$ decrease with increasing hexagonality. Between 3C and 2H, $B_{\text{VRH}}$ increases by 0.4\%, which means that the hexagonal polytypes are slightly more resistant to compression. In contrast, $G_{\text{VRH}}$, which is a measure for rigidity, decreases by 1\% from 3C to 2H. The Young's modulus $E$, a measure of the stiffness, also decreases by 0.7\%. Lastly, Poisson's ratio $\nu$ increases by almost 2.5\% for 2H. To summarize, among all polytypes, 3C is most easily compressible uniaxially but also has the highest rigidity and stiffness.

To gain more insight in the effect of strain on the polytype stability, we calculate the energy-strain curves for 3C and 4H. During growth, the vertical $[0001]$ direction can relax towards the vacuum (lattice constant $c$), whereas the  $[11\bar{2}0]$ direction (lattice constants $a$ and $b$) is prescribed by the substrate's lattice. Therefore, we consider biaxial strain by gradually changing lattice constants $a$ and $b$. The energy-strain curves are approximately parabolic with a minimum for the relaxed lattice parameters as observed in Section~\ref{ssec:bulk}. The seemingly identical curves are visualized in the inset of Fig.~\ref{fig:ev_curve}. Therefore, we show the difference in the energy-strain curves of 3C and 4H in the main plot. We choose these polytypes because 3C inclusions in 4H-SiC epitaxy is one of the main challenges during growth \cite{kimoto_bulk_2016, guo_understanding_2017, yamashita_characterization_2015}. (For a comparison of the energy-strain curves of 3C-2H and 4H-6H, see Ref. \onlinecite{kang_governing_2014}.) Additionally, we investigate the influence of the XC functional by comparing LDA, PBE, and PBEsol.

The difference in the energy-strain curves is plotted versus biaxial strain as a percentage of the lattice constants. In order to study the effect of solely the biaxial strain, we set the energy difference $E_{\text{4H}}-E_{\text{3C}}$ to zero at the relaxed lattice constants (0\% biaxial strain). All three functionals show positive energy differences at very high tensile and compressive strains $>\pm4\%$, indicating that 3C is stabilized by high stress and strain incidences. The results of the different XC functionals differ qualitatively for low strains. However, a 1\% strain is already very large; to illustrate, $\Delta a$ between the 3C and 4H lattice is -0.05\% \cite{levinshtein_properties_2001}. 

\begin{figure}[t]
	\includegraphics[width=\columnwidth]{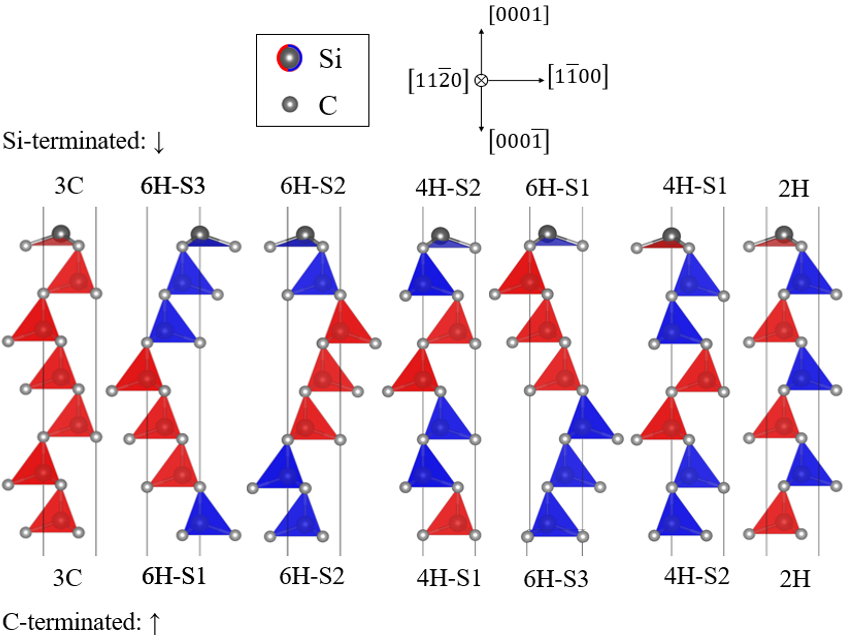}
	\centering
	\caption{The surface structures investigated in this work. The Si-terminated (C-terminated) surfaces are oriented towards the top (bottom). The hexagonality of the surfaces is shown in Table \ref{tab:surface}. Each unit cell has a 12 $\textrm{\AA}$ vacuum, one dangling bond and an H-atom bound to the (no longer) dangling bond of the opposite surface.}\label{fig:surface_schematic}
\end{figure}

PBE stabilizes 4H for moderate tensile and 3C for compressive strains. In contrast, LDA stabilizes 3C for tensile and compressive strains, the polytypes are predicted more or less equally stable until around -2\% strain. The PBEsol curve is approximately an average of the LDA and PBE curves, which could be explained by the fact that PBEsol was designed to solve the overbinding of LDA and underbinding of PBE \cite{perdew_restoring_2008}. For all functionals, 3C is stabilized by compression, which is in agreement with its lower bulk modulus $B$. We calculate the relative energy differences for the 2H and 6H polytype at -1\% strain and find values similar to 4H. To conclude, at realistic strains of $<1\%$, the energy differences are of the order of $\pm0.3$ meV/SiC. To obtain the full bulk thermodynamic stability of the polytypes, one should include the $T=0$~K internal energy (Sections~\ref{ssec:bulk} and~\ref{ssec:bulk_vdw}) and temperature contributions (Section~\ref{ssec:results_thermal}). Compared to the energy scale of other physical contributions, one can consider the elastic contribution negligible. 

\subsection{Surface energetics}
The surface properties play an important role during the epitaxial growth process of different polytypes, which commonly occurs along the [0001] direction as the industry standard. Along this crystal axes, SiC exhibits two polar surfaces, either Si- or C-terminated, corresponding to the (0001) and $(000\bar{1})$ surfaces, respectively. Growth experiments have produced evidence that differences in surface energy could be the governing factor in polytype stability \cite{stein_influence_1992}. Mercier and Nishizawa \cite{mercier_role_2012} calculated the surface energies with DFT employing the PW91 functional. They found that the relative stability of the Si- and the C-terminated surfaces is very different. Here, we study the surface energy of the various polytypes with conventional DFT methods and the DFT-D2/3 vdW correction.

\begin{table}[t]\caption{Overview of the surface structures. For 4H and 6H, there exist two and three different surface terminations, respectively. For each structure we list the bilayer stacking, ordering of h/k sites, the bulk- and surface hexagonality.}\label{tab:surface}
	\setlength{\tabcolsep}{8.5pt}
	\begin{tabular}{llll}
		\hline
		Polytype & Stacking                                                                            & \begin{tabular}[c]{@{}l@{}}Bulk\\ hexagonality\end{tabular} & \begin{tabular}[c]{@{}l@{}}Surface\\ hexagonality\end{tabular} \\ \hline
		3C         & \footnotesize{$\lvert$k)$_\infty$}    & 0   &         0               \\
		6H-S3     & \footnotesize{$\lvert$k$_2$k$_1$h)$_\infty$}   & 1/3 (33\%)   &  1/7 (14\%)    \\
		6H-S2     &    \footnotesize{$\lvert$k$_1$hk$_2$)$_\infty$}  &   1/3 (33\%)   &   2/7 (29\%)        \\
		4H-S2      &    \footnotesize{$\lvert$kh)$_\infty$}     &           1/2    (50\%)          &   1/3 (33\%)        \\
		6H-S1    &       \footnotesize{$\lvert$hk$_1$k$_2$)$_\infty$}  &   1/3 (33\%)        &   4/7 (57\%)   \\
		4H-S1      & \footnotesize{$\lvert$hk)$_\infty$ } & 1/2 (50\%) &                  2/3 (67\%)                       \\
		2H          &  \footnotesize{$\lvert$h)$_\infty$}   & 1 (100\%)   &               1 (100\%)                      \\ \hline
	\end{tabular}
\end{table}

In Fig.~\ref{fig:surface_schematic} and Table~\ref{tab:surface}, an overview of the calculated surface structures is shown. The 3C (2H) polytype have either only cubic (k) (hexagonal (h)) sites, which leads to only one surface structure. Whereas for 4H and 6H, the alternation of k- and h-sites leads to the occurrence of two and three surface-types per polytype, respectively. In total, we evaluate seven surface structures with different surface hexagonalities \cite{vignoles_iterations_1993, wang_surface_2010}. For each type, the Si- and C-terminated surfaces are treated separately. We construct our slab models by passivating the dangling bonds on the opposite side with hydrogen atoms. The surface energy of either the Si- or C-termination is calculated by 
\begin{equation}\label{eq:surface}
\sigma_{\text{Si/C}} = \frac{E_{\text{slab}}-n_{\text{SiC}}\mu_{\text{SiC}}-n_{\text{H}}\mu_{\text{H}}}{A}-\sigma_{\text{H}},
\end{equation}
where $E_{\text{slab}}$ is the total energy of the slab, $n_{\text{SiC}}$ and $n_{\text{H}}$ are the number of SiC-pairs and hydrogen atoms, $\mu_{\text{SiC}}$ and $\mu_{\text{H}}$ are the chemical potentials of one SiC-pair (derived from the bulk energy) and one hydrogen atom, $A$ the surface area, and $\sigma_{\text{H}}$ the energy of the opposite H-passivated surface. The H-passivated surface consists either of a Si-H or a C-H bond. $\sigma_{\text{H}}$ is approximated by calculating the average energy of the two H-passivated surfaces as
\begin{equation}\label{eq:h-surface}
\sigma_{\text{H}} = \frac{E_{\text{H-slab}}-n_{\text{SiC}}\mu_{\text{SiC}}-n_{\text{H}}\mu_{\text{H}}}{2A},
\end{equation}
where $E_{\text{H-slab}}$ corresponds to the energy of a slab with a H-atom on both sides. Through the combination of equations \eqref{eq:surface} ($n_{\text{H}} = 1$) and \eqref{eq:h-surface} ($n_{\text{H}} = 2$), $\mu_{\text{H}}$ can be eliminated, which allows us to calculate the absolute surface energies. 

\begin{figure}[t]
	\includegraphics[width=\columnwidth]{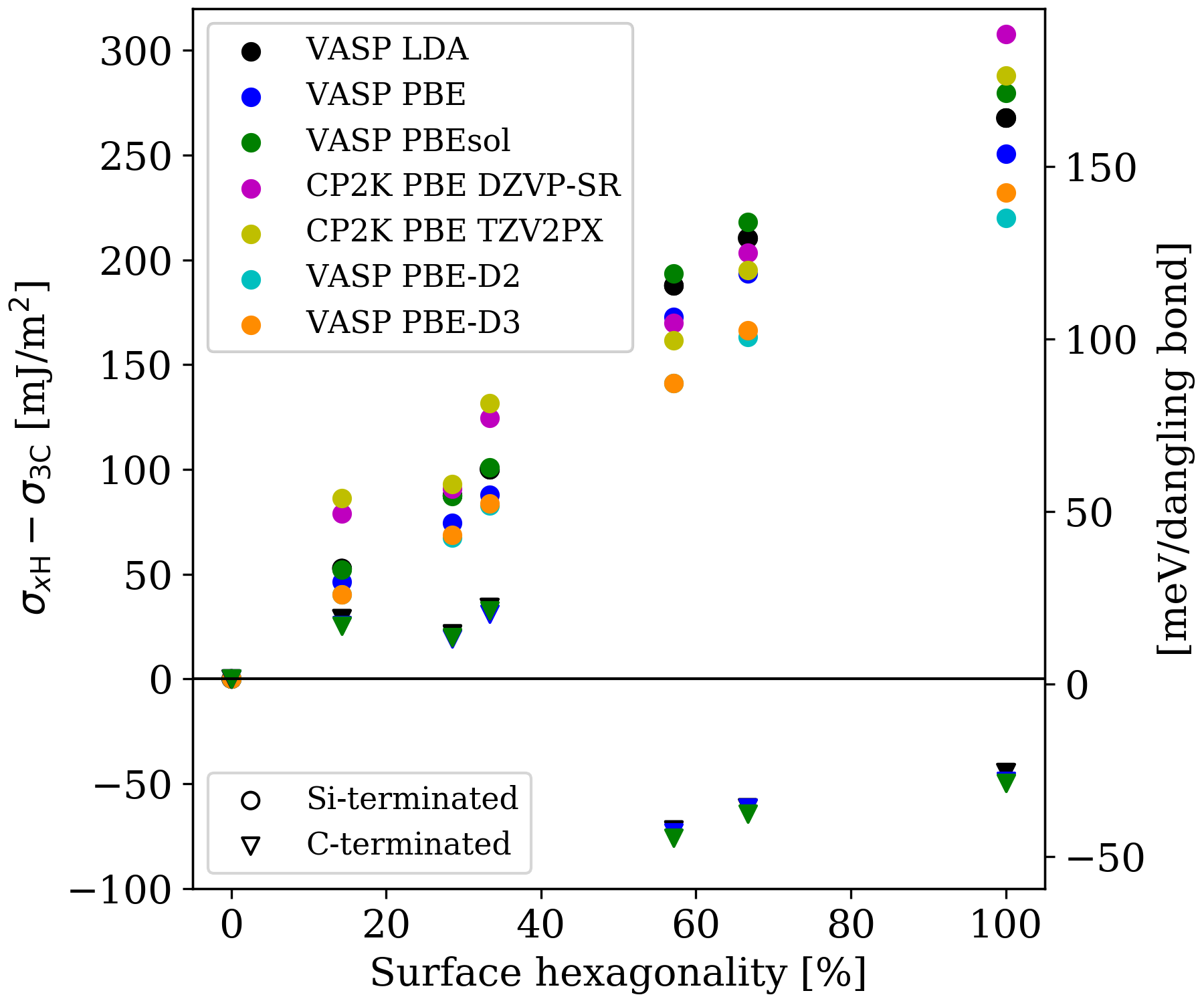}
	\centering
	\caption{Surface energy relative to the 3C polytype versus surface hexagonality, which is determined by the polytype and the termination of the surface.} \label{fig:surface}
\end{figure}

Previous calculations of the bulk energies have shown the importance of benchmarking the different available DFT methods. In Fig.~\ref{fig:surface}, we show the relative surface energies for six different methods in the same way as Ref. \onlinecite{mercier_role_2012}. The energies of the $x$H surfaces relative to 3C are plotted versus surface hexagonality. For all evaluated methods, the energy of the Si-terminated surfaces is linearly proportional to the surface hexagonality. We investigate the effect of the XC functional with VASP. The difference between LDA, PBE and PBEsol is small compared to the energy differences among the surface structures. Similar to the bulk calculations, LDA and PBEsol result in very similar relative energies, which are slightly higher compared to PBE. Although we include a dipole correction, its effect is negligible on the scale of the surface energy differences.

The discrepancy between the VASP-PBE and CP2K-PBE methods is small but notable, up to 18 meV/dangling bond for the low hexagonality surfaces. A possible explanation could be that CP2K-GPW and VASP-PAW treat the vacuum differently due to their different representation of the wavefunction. The vdW dispersion correction methods DFT-D2/3 lower the energy of the hexagonal surfaces slightly. 

The different DFT methods only show small quantitative differences, but the overall trend is retained. For the Si-face, we see a linear trend and the 3C surface is most stable by up to 158 meV/dangling bond (2H). The C-terminated surfaces show a different ordering in surface stability. The high hexagonalities 6H-S1, 4H-S1, and 2H are more stable by up to 17 meV/dangling bond, whereas the low hexagonalities 6H-S3, 6H-S2, and 4H-S2 are less stable than the 3C surface by up to -42 meV/dangling bond. These calculations were performed without considering spin-polarization and surface reconstruction. Since all DFT methods result in a consistent trend, we continue to investigate the surface properties with VASP-PBE. 

\begin{figure}[t]
	\includegraphics[width=0.9\columnwidth]{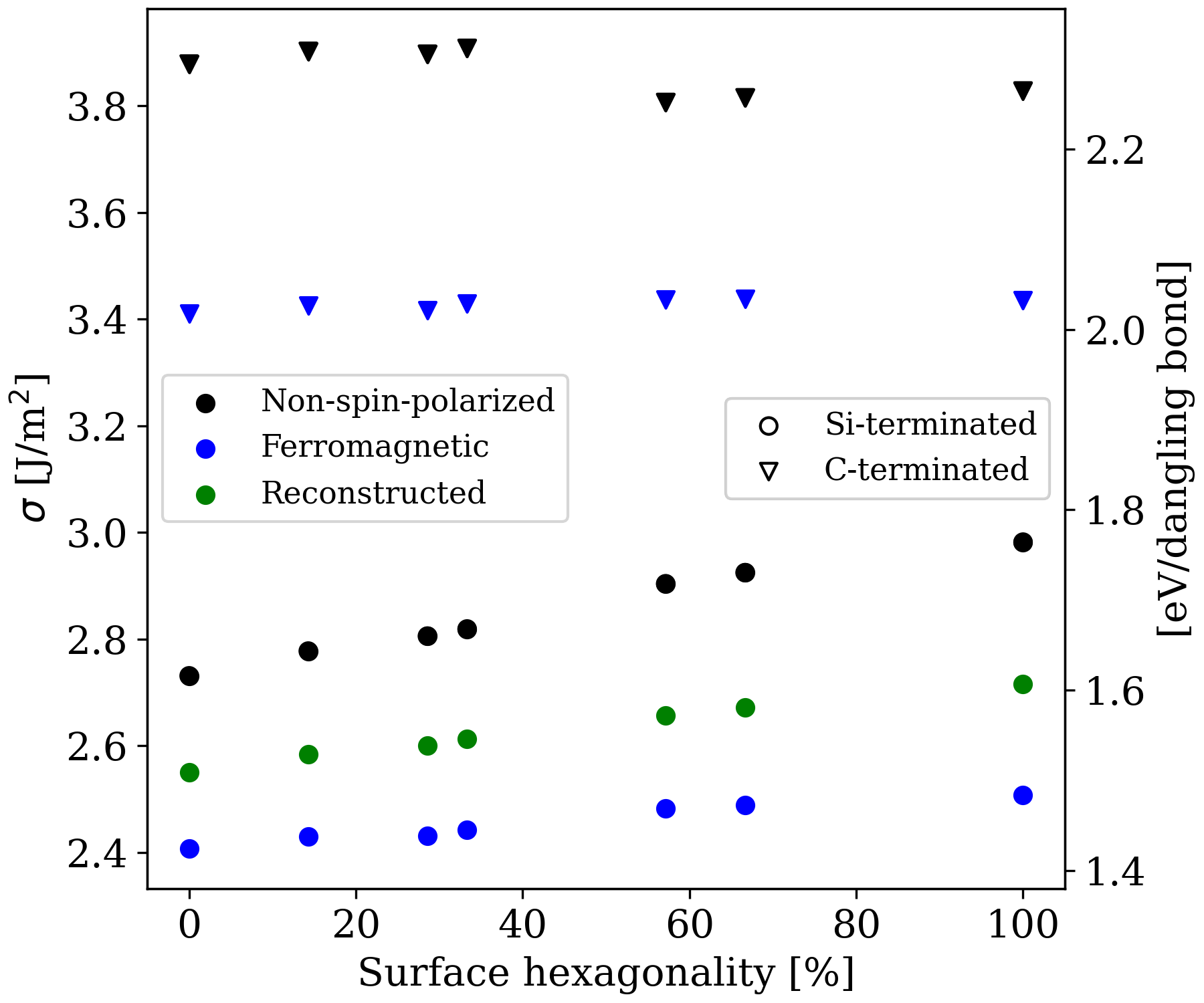}
	\centering
	\caption{Absolute energies of the Si- and C-terminated surfaces versus hexagonality for the non-spin-polarized and ferromagnetic ideal surface, and, in the case of Si-termination, a $2\times1$ (non-magnetic) reconstruction. All results are obtained with VASP-PBE.} \label{fig:surface_mag}
\end{figure}

The absolute energies of the Si- and C-terminated surfaces are plotted in Fig. \ref{fig:surface_mag} and tabulated in Table~IV of the SM \cite{SM}. The non-spin-polarized (black) values coincide with the relative energies in Fig. \ref{fig:surface}. Generally, the Si-face has lower energy than the C-face. Without spin-polarization, the Si-terminated surface undergoes a $2\times1$ buckled reconstruction, which has been previously reported and is shown in Fig. 1 of Ref. \onlinecite{seino_microscopic_2021}. Through the reconstruction, the surface energy is lowered between 106 (3C) and 160 (2H) meV/dangling bond. Even though the reconstruction is favorable, it has not been observed experimentally \cite{fissel_si_2003, coati_3$times$_1999, northrup_theory_1995}. 

The discrepancy between DFT and experiment is elucidated by introducing spin-polarization, with which we find a new ferromagnetic ground state for both the Si- and C-face. In this case, the electron of the surface dangling bond fully occupies one spin channel, which lowers its energy significantly. The reduction is between 191 (3C) and 281 (2H) meV/dangling bond for the Si-terminated and between 219 (6H-S1) and 284 (6H-S2) meV/dangling bond for the C-terminated surfaces. Note that for this ferromagnetic state the buckled structure is not a local minimum on the potential energy surface and its relaxation leads back to the pristine, non-reconstructed surface. We also consider some of the smallest periodicity antiferromagnetic patterns and found that they produce near to identical values as the ferromagnetic surface. The spins of the dangling bond are uncorrelated and the ground state is achieved as long as the dangling bonds fully occupy one of the spin channels. 

The spin-polarization reduces the energy differences between the polytypes. For the Si-face, the linear trend is retained but the largest energy difference (between 3C and 2H) is now 59 instead of 158 meV/dangling bond. For the C-face, the 3C surface is now most stable and the energy differences are between 4 and 16 meV/dangling bond compared to the hexagonal polytypes. The energy differences per unit cell (with one surface dangling bond) are more significant than those of the bulk thermodynamic stability for growth on the Si-face.

\section{Conclusion}
In summary, we have systematically evaluated the energetics of the 3C, 2H, 4H, and 6H SiC polytypes for various attributes that play a role during epitaxial growth. We investigated the thermodynamic stability of bulk phases, both at zero and elevated temperature, and under lateral biaxial strain. Furthermore, we computed the surface energetics of the (0001) Si- and (000$\bar{1}$) C-terminated surfaces. 

Since the bulk energy differences at $T=0$~K between the polytypes are very small, of the order of 1 meV/SiC. We carefully studied the effect of different DFT implementations and XC functionals. The results obtained using the VASP, CP2K, and FHI-aims codes are consistent when using the PBE functional. The accurate hybrid functional HSE06 gives results consistent with PBE, indicating that PBE describes the SiC polytypes adequately. LDA and PBEsol predict the $x$H polytypes to be around -1 meV more stable than those obtained with PBE. PBEsol is especially suitable for the geometrical properties, matching the experimental lattice parameters and solving the well-known under- and overbinding of PBE and LDA, respectively. The two meta-GGA functionals, SCAN and RTPSS, deviate by about 0.5 to 1 meV/SiC from PBE, but with an opposite trend. 

In addition, we have evaluated nine vdW methods, which are notably more inconsistent in predicting the polytype stability than the non-vdW methods. The DFT-D2 and the vdW-functionals show a significant correction to the PBE energies resulting in a greater stability of the 3C phase than the hexagonal polytypes. However, the lattice ratios predicted by these methods deviate more severely from experimental values, which could be an indicator that these methods are less suited for the SiC-system. The most advanced vdW methods have a negligible energy correction and predict the lattice parameters correctly like the conventional DFT methods. Thus, in contrast to Ref.~\onlinecite{kawanishi_effect_2016} and Ref.~\onlinecite{scalise_temperature-dependent_2019}, we conclude that the vdW correction methods introduce spurious errors when computing energetics of the bulk SiC system, except for the most advanced vdW methods.

\begin{figure}[t]
	\includegraphics[width=\columnwidth]{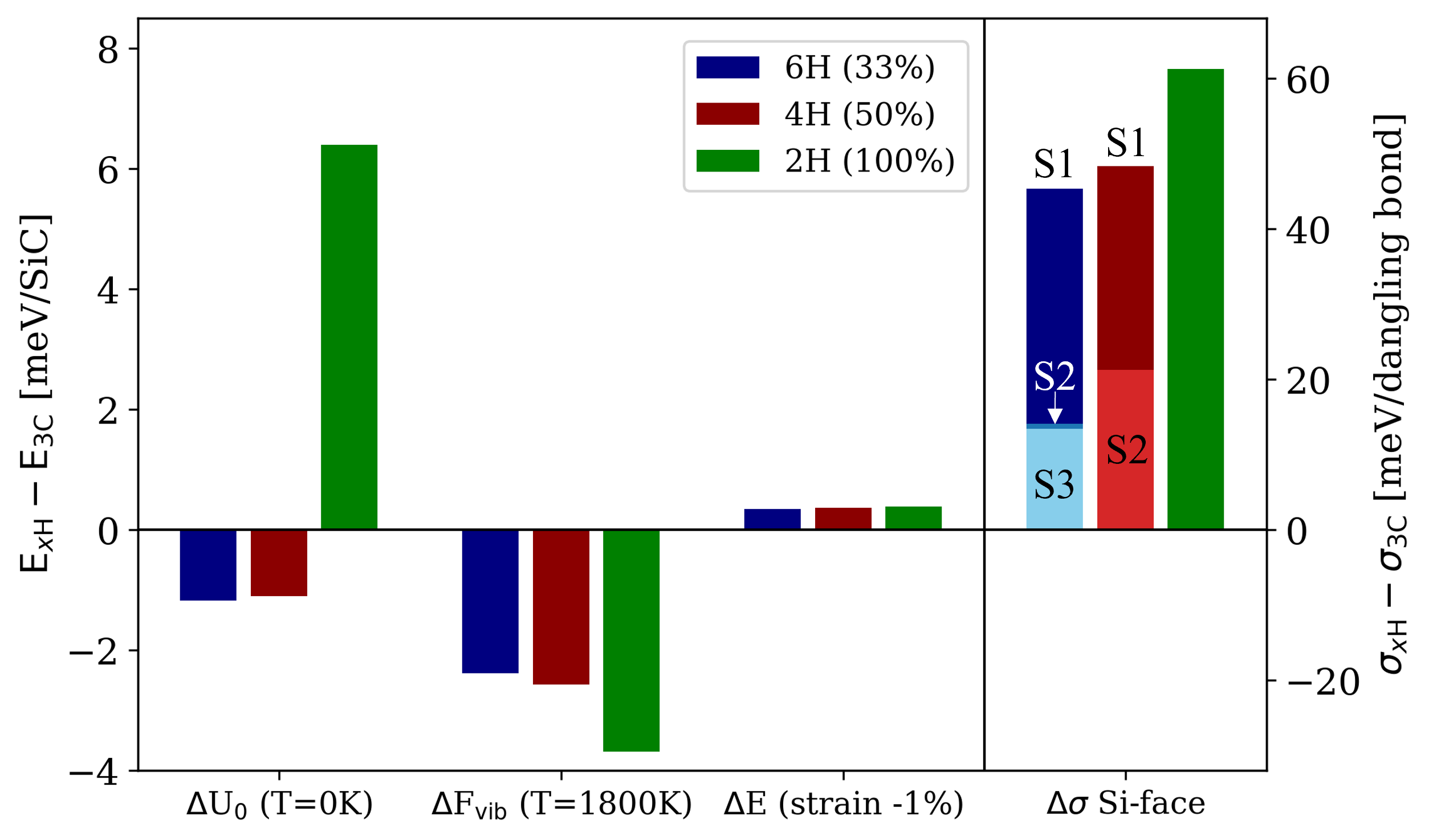}
	\centering
	\caption{Comparison of the energy contributions studied in this work: the bulk internal energy $\Delta U_0$ at $T=0$~K, the phonon vibrational energy $F_{\mathrm{vib}}$ at $T=1800$ K, the elastic contribution $\Delta E$ at 1\%, and the Si-face surface energy. The differences between 3C and $x$H polytypes are plotted for each contribution based on the results obtained by the VASP-PAW-PBE method. The surface energy differences are labelled according to Table \ref{tab:surface}.}\label{fig:conclusion}
\end{figure}

To compare the various factors that affect polytype stability, we show a comparison of energy contributions in Fig. \ref{fig:conclusion}. At 0 K, the 3C, 4H and 6H polytypes have very similar energies with 4H and 6H being -1 meV/SiC more stable than 3C. The 2H polytype is in contrast around 5 meV/SiC less stable than 3C. The temperature contribution $\Delta F_{\mathrm{vib}}|_{T=1800\mathrm{K}}$ stabilizes the hexagonal polytypes by about -3 meV/SiC for 4H and 6H, and -4 meV/SiC for 2H. Contrary to Scalise et al.  \cite{scalise_temperature-dependent_2019}, whose relative energies at 0 K were based on the spurious vdW PBE-D2 methods, our results show that the bulk phase of 4H and 6H are always more stable than 3C. The effect of biaxial strain is negligible compared to the other bulk contributions for realistic strains of up to 1\%, in contrast to the reports of Kang et al.\cite{kang_governing_2014}

Since the bulk thermodynamic stability cannot explain why 3C growth is observed at low temperatures and why 3C inclusions are found during epitaxy, we also considered the surface energetics. For the Si-terminated surfaces (commonly exposed during epitaxial growth), the surface energy of 3C is considerably lower than that of the hexagonal polytypes. The low-hexagonality surfaces 6H-S3 (14\%), 6H-S2 (28\%), and 4H-S2 (33\%) exhibit smaller energy differences of up to 20 meV/dangling bond. For the 6H-S1, 4H-S1, and 2H surfaces with higher hexagonality, the energy differences are between 45 and 59 meV/dangling bond. During 4H (6H) SiC epitaxial growth in the [0001] direction, the 4H-S1 and 4H-S2 (6H-S1, 6H-S2, and 6H-S3) surfaces alternate and are thus both of importance. 

To conclude, it seems that the difference in surface energy is likely the driving force for 3C-nucleation, even though the difference in bulk energy slightly favors the 4H and 6H polytypes. It is crucial to take into account all contributing factors to understand the intricate mechanisms that govern the SiC epitaxial growth.

\begin{acknowledgments}
We would like to thank Bernd Thomas, Daniel Baierhofer, Benjamin Marchetti and Dick Scholten from Robert Bosch GmbH for insightful discussions on SiC epitaxial growth.
\end{acknowledgments}

\bibliography{paper_ramakers_resub2}

\end{document}